\documentstyle[aps,epsfig,multicol]{revtex} 
\begin{document}
\draft
\title{Quantum Phase Transitions in the  Ising model in  spatially modulated 
 field}
%\author{Parongama Sen\\
%\author{Parongama Sen}
%
%\address
%Department of Physics, University of Calcutta,\\ 
%92 A.P. C. Road, Calcutta 700009, India.\\
%e-mail paro@cucc.ernet.in}
%%\author{Parongama Sen\\
\author{Parongama Sen}

\address{Department of Physics, University of Calcutta,
92 A.P. C. Road, Calcutta 700009, India.\\
e-mail paro@cucc.ernet.in}
\maketitle

\maketitle

\begin{abstract}

The phase transitions in the transverse field Ising  model in a competing 
spatially modulated (periodic and oscillatory) longitudinal field are  studied 
numerically. 
There is a multiphase point in absence of the transverse field where 
the  degeneracy  for a longitudinal field of  wavelength
$\lambda$ is $(\frac {1 + \sqrt{5}}{2})^{2N/\lambda}$
for a system with $N$ spins, an exact  result obtained from the known 
result for $\lambda =2$.
The   phase transitions in the $\Gamma $ (transverse field) versus  $h_0$ 
(amplitude of the longitudinal field)  phase diagram are obtained from the 
vanishing  of the mass gap $\Delta$.
We find that for all the phase transition points
obtained in this way, $\Delta $ shows finite size scaling behaviour 
signifying a continuous phase transition everywhere.
The values  of the  
%dynamic exponent  and the correlation length
critical exponents  show that the model belongs to the  universality 
class of the two dimensional  Ising  model.
The longitudinal  field is found to 
have the same scaling behaviour as that of the transverse field, which seems 
to be a unique feature for the competing field. 
The phase boundaries for two different wavelengths of the modulated
field are obtained.   
Close to the  multiphase point at $h_c$, the phase boundary behaves as  
$(h_c - h_0)^b$, where $b$ is also $\lambda$ dependent. \\\


\end{abstract}
%\twocolumn
\begin{multicols}{2}    

\section{Introduction}
Competitive interactions in magnetic models can generate  
frustration leading to the  existence of a large number of degenerate states
\cite{Liebemann}.
For example, in the ANNNI (axial next nearest neigbour Ising) 
model \cite{Liebemann,elliot,selke}, 
 the presence of a second neighbour 
 antiferromagnetic interaction induces a competition. 
%This  competition  can generate a frustration when the
%strength of the next neighbour interaction is exactly half of that of the
%nearest neighbour. 
In absence of any fluctuation, a  large number of degenerate states 
 exist at a  so called  multiphase point. 
In spin glasses \cite{Binder}, interactions are random, and
the mixture of ferromagnetic and antiferromagnetic interactions result in  
frustration and huge degeneracy.

Phase transitions, both classical and quantum, in many of these  
systems 
are well studied \cite{Liebemann,elliot,selke,Binder,CDS,sachdev}. 
While the short range Ising like models
do not have any finite temperature phase transition in one dimension,  
the phase diagram
in higher dimensions often shows very rich structures. 
%various modulated 
%phases in three dimensions and Lifshitz points in the ANNNI model. 
The quantum phase transitions  are  present  even in one dimension.
It is interesting to observe that the  zeroth order thermal fluctuation in two
dimensions and correspondingly zeroth order quantum fluctuations in 
one dimension can destroy all order at the fully frustrated multiphase
 point \cite{selke,CDS} in ANNNI-like models.

The Ising models in transverse field form a special class of 
quantum systems and to the best of our knowledge, have
different features in comparison to quantum spin models in which the 
quantum effect comes through the cooperative interaction.   
 
In this paper, we will discuss  a quantum Ising 
model with competition,
where the competition is generated by the presence of an external
longitudinal  
 field which is spatially periodic. We choose the 
periodicity of the field to be commensurate with the lattice periodicity.
The field is also oscillatory so that the net field on the 
entire system sums out to zero. Classically, the ferromagnetic Ising system 
 will not undergo any phase transition in a uniform field.  On the other hand, 
spatially modulated 
fields can play important roles in critical phenomena. Classical systems
in periodic potential (Frenkel-Kontorova model), for example, 
show very interesting features leading to novel concepts like the  
devil's staircase  \cite{Aubry}.

Some interesting and relevant features of the  model in the classical limit
are discussed in section II.
The   phase transitions  in the ferromagnetic model in the  modulated 
field in one dimension
at zero temperature   driven by quantum fluctuations
generated by  the  transverse field are described  in section III.  
The results are discussed and  concluding  remarks are made in section IV.

\section{The  model: Classical ground state and degeneracy}

The Hamiltonian for the system with $N$ spins  is given by

\begin{equation}
H_{q} = -J\sum _{i=1}^{N}S^{z}_{i}S^{z}_{i+1} -\sum_{i=1}^{N}h_iS^{z}_{i}.
-\Gamma \sum_{i=1}^N S^{x}_i.
\end{equation}

The form of the longitudinal periodic field  $h_i $ is like this: $h_i = h_0$ at $i=(n\lambda +1) $ to 
$(2n+1)\lambda/2$ ($n=0,1,....,N/\lambda -1)$ and $-h_0$ elsewhere. 
 The  wavelength of the field is 
 denoted by $\lambda$. $\Gamma$ is the strength of the transverse field.

Note that the modulated field can be chosen 
to be of various  forms, for simplicity we will choose a square wave form.
While the ferromagnetic interaction tries to  align the
spins parallely, the periodic  field will try to  modulate the system
spatially.  We will consider periodic boundary condition.           

We will first discuss some known results in the 
context of our model in the classical limit (i.e., $\Gamma = 0$). 
The classical Hamiltonian  is 
\begin{equation}
H_{cl} = -J\sum _{i=1}^{N}S^{z}_{i}S^{z}_{i+1} -\sum_{i=1}^{N}h_iS^{z}_{i} .
\end{equation}

Let us first take the case  $\lambda =2$: the field is 
positive on odd sites and negative on even. Now one can simply 
use a transformation $S^{z}_{2i} \rightarrow -S^{z}_{2i}$. This 
effectively makes the Hamiltonian 
\begin{equation}
H_{cl} = J\sum _{i=1}^{N}S^{z}_{i}S^{z}_{i+1} -h_0\sum_{i=1}^{N}S^{z}_{i} ,
\end{equation}
i.e, an  antiferromagnetic system in a uniform field \cite{domb,aharony} which
is solved exactly by transfer matrix method.

Again, if we consider the ANNNI chain, the Hamiltonian is:
\begin{equation}
H_{ANNNI} = -J_1\sum _{i=1}^{N}S^{z}_{i}S^{z}_{i+1} 
 +J_2\sum _{i=1}^{N}S^{z}_{i}S^{z}_{i+2}.
\end{equation}
The  above can be cast to the form of (3) by the transformation $S^{z}_{i}
S^z_{i+1} \rightarrow S^z_i$ \cite{Liebemann} with $J_1$ playing the role of 
$h_0$
and $J_2$ the role of $J$. Note that $J_2$ is positive, describing 
antiferromagnetic interaction by definition. The sign of $J_1$ is
unimportant.

The two  models described by  equations (3) and (4) are well 
studied \cite{domb,Liebemann,selke}  and they have 
similar classical ground states: upto $J/h_0 = 0.5 $ (or $J_2/J_1 = 0.5$),
the system is antiferromagnetic (ferromagnetic for  ANNNI when $J_1 >0$), and 
paramagnetic (modulated for ANNNI) beyond this value. An infinite number
of degenerate states exist at this point. The degeneracy can be calculated
exactly \cite{Liebemann,domb} and is equal to $\tau^N$ where 
$\tau = (1+ \sqrt{5})/2$. 
In the language of the
original classical model defined in (2) with $\lambda = 2$, one has 
ferromagnetic 
phase upto $h_0/J = 2$ and  paramagnetic phase for $h_0/J > 2$.

Even when $\lambda \neq 2$, one can rescale the system by a factor of
$\lambda/2$ so that the Hamiltonian  is effectively  written as in (3).
It can easily be checked that the multiphase point is 
now at $h_c/J = 4/ \lambda$
and the degeneracy given by $\tau^{2N/\lambda}$.

 It is extremely important to note here that although the antiferromagnetic and
ferromagnetic systems  have identical  critical behaviour,
the presence of a field  makes things drastically different. Therefore  the 
modulated field in the ferromagnetic system is effectively a 
uniform field in the mapped antiferromagnetic system. 
We  will specifically take the  interaction to be ferromagnetic in order 
to avoid any ambiguity or confusion.
%If, on the other hand, 
%in (1), we have an antiferromagnetic system to begin with, then 
%for $\lambda =2$, there is no competition: however, for other values of
%$\lambda$, competitive effects will appear,
%the nature of which maybe different from that of the ferromagnetic model (e.g.,
%one may have a different $\lambda$ for the equivalent 
%ferromagnetic model).

\section{Quantum Phase Transition and Phase diagram at T=0}

There are two limits in which  the results are exactly known for the 
quantum Hamiltonian (1) at zero temperature. 
The classical limit $\Gamma = 0$ is already discussed.
In the limit $h_0 = 0$, the problem is also exactly  solved \cite{Pfeuty}: 
there is a phase transition at a finite value of the transverse 
field: $\Gamma_c(0)/J =  1$. This is a quantum critical point and the
phase transition is continuous in nature.
In a spatially periodic and oscillatory 
longitudinal   field which is competing in nature,  
the phase transition should
 occur even at lower 
values of the transverse field, $\Gamma_c (h_0)$.

Vanishing of the mass gap $\Delta (= E_1 -E_0$ where $E_1$ and $E_0$
are the energies of the first excited and the ground state respectively)
 in the thermodynamic limit is a signature  of 
 a continuous phase transition.
We obtain the mass gaps for finite chains of length $L$ ($N=L^d = L$, where
$d$ the dimensionality is one  here).

 We employ Lanczos' method to diagonalise the Hamiltonian 
matrix which is calculated in the basis in which $S^{z}$ is diagonal.
The field being spatially modulated and also commensurate 
with the lattice periodicity,  the system sizes are restricted
 to specific values
for a given wavelength. On the other hand, one requires
data for different   system sizes for a finite size scaling procedure.
Since it is difficult to work with large
sizes in a diagonalisation scheme anyway, we have obtained the phase 
diagrams for 
$\lambda$ = 2 and 4 only. The maximum size taken for both cases is $L=16$.
The result for a very large value of
$\lambda$ can be easily guessed. 

We indeed find that $\Delta$ vanishes with $1/L$ not only for 
small values of $h_0$ but right upto the critical field $h_c = 4J/\lambda$.
The critical points $\Gamma_c(h_0)$ can be obtained by extrapolation, and a 
more accurate  estimation is done using finite size scaling of the mass gap,
which  shows scaling behaviour. We defer a detailed discussion
on the scaling of mass gaps and nature of transitions and first 
summarise the results for the phase diagram.

In Fig. 1, the phase diagrams for $\lambda =2$ and 4 are shown.
We have  calculated the ferromagnetic order: 
$\langle m\rangle  
=\frac {1} {L} 
 \sum\langle S^z\rangle $ 
as well as the "staggered magnetisation"
defined as $\frac {1} {L} \sum \langle S^{z}_ih_i\rangle /h_0$.
 The phase boundary we  obtain separates a ferromagnetic phase with
nonzero $\langle m \rangle$ and a 
paramagnetic phase with $\langle m \rangle = 0$. However, 
since the longitudinal field is always present, 
the staggered magnetisation is nonzero all over the   paramagnetic phase. 
% Only for $\Gamma/J  >  1.0$  and  $h=0$ one obtains a state with
%no order in agreement with the exact result.

It is also observed that 
as soon   as $h_c/J =4/\lambda $ is reached, the system behaves as if 
 only a field is present and therefore no phase transition  
is observed here for any non-zero $\Gamma$: the mass gap remains 
finite in the thermodynamic limit
almost without any  system size dependence.
As in some other models with competing interactions, in this model also, 
the zeroth order  quantum fluctuation
destroys the order at $h_c$ and  lifts the degeneracy.

The entire phase boundary  could not be fit to any simple
form;  we could  obtain the behaviour close to the  
 multiphase point
quite accurately for both $\lambda = 2$ and $\lambda = 4$. Here the 
phase boundary  appears 
to be of the form $(4/\lambda - h_0/J)^b$ where $b = 0.77 \pm 0.02$ for $\lambda =2$
and $b = 0.49  \pm 0.01$ for $\lambda =4$. 

For $\lambda \rightarrow \infty$,
the critical field $h_c \rightarrow 0$, the ferromagnetic phase exists 
only for $h_0 =0$, and the phase boundary will 
coincide with the $h_0 = 0$ axis. The exponent $b$  decreases with 
$\lambda$, ultimately vanishing for $\lambda \rightarrow \infty$.  

We end this section with a  discussion on the scaling behaviour of 
mass gaps and the nature of phase transitions.
In zero longitudinal field,  the finite size scaling for the
mass gap gives:
\begin{equation}
\label{massgap}
\Delta \sim L^{-z} f((\Gamma-\Gamma_c)  L^{1/\nu})
\end{equation}
where $z$ is the dynamic exponent and $\nu$ is the 
correlation length exponent.
$z=1$ and $\nu = 1$ for the one dimensional 
transverse field Ising model. The latter 
can be mapped to a two dimensional classical Ising model.
 $f$ is a universal scaling function.  

We find the above finite size scaling form to be valid for non-zero values of 
$h_0$    
in the sense  that when $L\Delta$ is plotted 
against $(\Gamma -\Gamma_c(h_0))L^{1/\nu}$ the data for different system sizes
again  collapse for any  value of  $h_0$ (see Fig 2). 
The data collapses are obtained with  $z = 1$ and $\nu = 1$
for all $h_0$, indicating  that the model belongs to the  
universality class of the two dimensional
classical Ising model even for non-zero longitudinal modulated fields.  
As $h_0$ is increased, the collapse gets limited to a smaller 
region.  The scaling function  depends on the value of $h_0$.

The phase transition points depend on the values of both $\Gamma$ and $h_0$.
	The plots in Fig. 2 are done by keeping $h_0$ fixed and varying 
$\Gamma$ around the critical value. One can study the scaling  the 
other way also: keep $\Gamma$ fixed and vary $h_0$. We find that there is a 
collapse of data for different sizes again, as shown in Fig. 3.
The striking feature is that the scaling argument is 
$(h_0-h_c(\Gamma ))L^{1/\nu}$
with $\nu=1$. This shows that the competing longitudinal field
scales in the same way as does the transverse field.
The results for $\lambda =4$  show the same scaling behaviour
as in Figs 2 and 3.
We discuss this feature again in section IV.

One may  conclude from the above observations 
that the phase transitions are continuous everywhere (except at $\Gamma =0$)
as the mass gaps vanish continuously and show scaling behaviour
as in conventional critical phenomenon. 
%There is no signature of a tricritical point anywhere.

\section{Discussions and Conclusions}  

The Ising model in the presence of a transverse field and a longitudinal
field cannot be 
solved exactly. Recently, most of the 
studies on transverse Ising models have involved 
randomness in the transverse  field,   interactions or longitudinal field.
We have considered a very simple model in which there are both quantum effects
and frustration, where the frustration comes as a result of
the modulated nature of the external field and no randomness is involved. 

In absence of the transverse field, the system has a multiphase point.
The  degeneracy at 
the multiphase point in the absence of fluctuations
is easily calculated for any $\lambda$ from the known result corresponding
to  $\lambda =2$.

 We have obtained the phase diagram of the model in the $\Gamma -h_0$ plane  
at zero temperature using numerical methods. 
Our results show  that a continuous phase transition occurs everywhere
except at the multiphase point where $\Gamma_c =0$ and a first order
phase transition is known to exist. The transition at this  point is of first
 order
in the sense that the order parameter (magnetisation) discontinuously
vanishes at this point. 
The values of the critical exponent $z$ and $\nu$ obtained in this model
are identical to those of the transverse Ising model, showing that the
periodic longitudinal field is irrelevant.
Thus it belongs to the classical two-dimensional Ising universality class.

Our conclusions  are based on the scaling behaviour of the mass gap.
As in conventional critical phenomena, it shows finite size scaling behaviour
close to the critical point.
It is to be noted that according to scaling arguments in classical
critical phenomena,
there are only two relevant fields for magnetic systems,  
temperature ($T$) and  magnetic field ($h$), and the
correlation length scales as:
\begin{equation}
\label{scale}
\xi = L ~ g((L^{1/\nu}t, hL^{(\beta+\gamma)/\nu}).
\end{equation}
The deviation from the critical temperature is denoted by $t$, $\beta$
 and $\gamma$ are the critical exponents associated with the 
order parameter and susceptibility respectively and $g$ is a universal scaling 
function.
Here  the field $h$ and the temperature have different scaling dimensions. 
There is, however, no criticality associated with the uniform ordering field
$h$.
In the present model, the two important quantities are
the transverse field and the competing longitudinal field.
The quantum fluctuations through $\Gamma$ may be compared
to the thermal fluctuations. The roles  of $\Gamma$ and 
$T$  may be considered  to be equivalent; however, the roles
of the competing field and ordering field are completely different. 
That is why a scaling form as in (\ref{scale}) will not be valid here. 
On the other hand, from the study of the 
finite size scaling of the mass gaps (section III), it appears
that the competing field and the transverse field have the same scaling 
behaviour.  We claim this to be a unique feature of the non-ordering 
competing field.

Beyond the multiphase point, the field is dominating,
and no  continuous phase transition  can exist here;  
the mass gap remains finite in  the thermodynamic limit.   
This is in contrast to the competitive models like ANNNI model,
where one obtains phase transitions beyond the fully frustrated 
point as well. This difference is of course due to the different roles
played by interaction and field.
The correspondence between the Ising model in competing field and
the ANNNI model mentioned in section II 
is valid only in one dimension and in the classical limit. 

From the present study, however, existence of other modulated  phases, 
separated by first order lines, cannot be ruled out. 
Such a possibility is strongest close to the multiphase point.
A closer inspection, eg., using perturbation methods could be 
done  from which it may also be possible to verify the behaviour
of the phase boundary and get an estimate of $b$ defined in section III.

A few other points need to be mentioned. In the random field 
transverse model, a tricritical point was obtained \cite{amit}.  
The phase diagram
of certain antiferromagnets like  
 $\rm{FeCl_2}$    
in a uniform field also show a 
tricritical point. In comparison, our model does not have any such point. 
Rather,   the
phase boundary  to the left of the multiphase point 
is very similar to that of the 
 ANNNI model in higher dimensions. 
In the latter also, the next neighbour interaction cannot change the 
universality class of the Ising model. 
The mapping to the antiferromagnetic model in the classical limit
and subsequent description of the ground state assumes that we are considering 
hypercubic lattices only, where, in absence of any field at zero temperature, the antiferromagnet  will 
order. This is not true in lattices in which the  
antiferromagnetic system   is geometrically frustrated 
(e.g., in a triangular lattice); such systems in  transverse 
 (and longitudinal) field
 in two dimensions have been recently studied \cite{sondhi}.

We have considered only a square wave form for 
the modulated field in this discrete model. 
The features should not be different for a different 
form, however, there will be  quantitative changes. 
In a continuum model, it may be more convenient to choose
a sinusoidal form.
More complex spin patterns are obtained in (antiferromagnetic) systems with 
long range interactions in a   field, e.g., the  long range 
antiferromagnet in a uniform 
field  \cite{Bak} exhibits a complete devil's staircase.
It is expected that there will also be very interesting features if the 
modulation in the field
 is incommensurate with the lattice periodicity.  \\\\

The author  is grateful to S. Dasgupta, A.Dutta and B. K. Chakrabarti for
very useful and enlightening discussions  and suggestions. She also thanks I. Bose for comments.

%\pagebreak

%\begin{references}  

%\end{references} 

%\pagebreak

\narrowtext
%\vskip -5.0cm
\begin{figure}
\caption{The phase boundaries between ferromagnetic and paramegnetic regions
are shown for the transverse Ising model in a spatially modulated 
field for two wavelengths $\lambda = 2 $ and $4$. 
Near $h_c(\lambda)/J = 
4/(\lambda)$, 
the boundaries  fit to the
form $(h_c(\lambda) -h_0)^b$, 
shown by the continuous curves, 
where $b = 0.77\pm 0.02 $ and $0.49 \pm 0.01$ for
$\lambda = 2 $ and $\lambda =4$ respectively.}

\caption{The scaling plot for the scaled mass gaps $L^z \Delta $ 
with $(\Gamma-\Gamma_c(h_0))L^{1/\nu}$ for $\lambda =2$. 
The four sets correspond
to the values of $h_0 = 1.9, 1.8, 1.6$ and $ 1.0$ from top to bottom, 
each showing the collapse of data points for $L = 8,10,12$ and $16$
with  $z=1$ and $\nu =1 $.}

\medskip
\caption{The scaling plot for the scaled mass gaps $L^z \Delta $ 
with $(h_0-h_c(\Gamma ))L^{1/\nu}$ for $\lambda =2$. 
The four sets correspond
to the values of $\Gamma  = 0.25, 0.45, 0.65$ and $ 0.95$ from top to bottom, 
each showing the collapse of data points for $L = 8,10,12$ and $16$
with  $z=1$ and $\nu =1 $. }
\end{figure}

\end{multicols}  
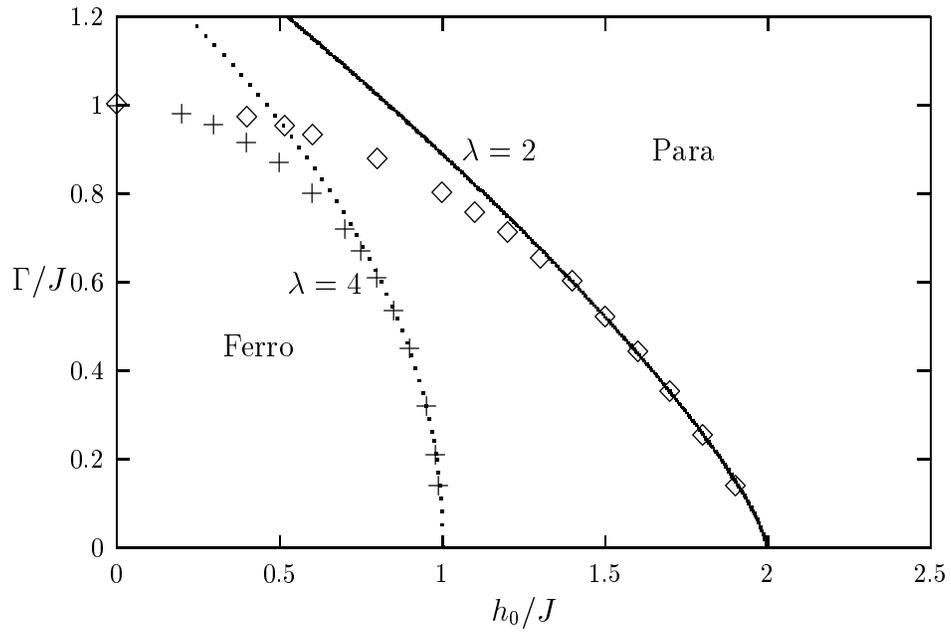
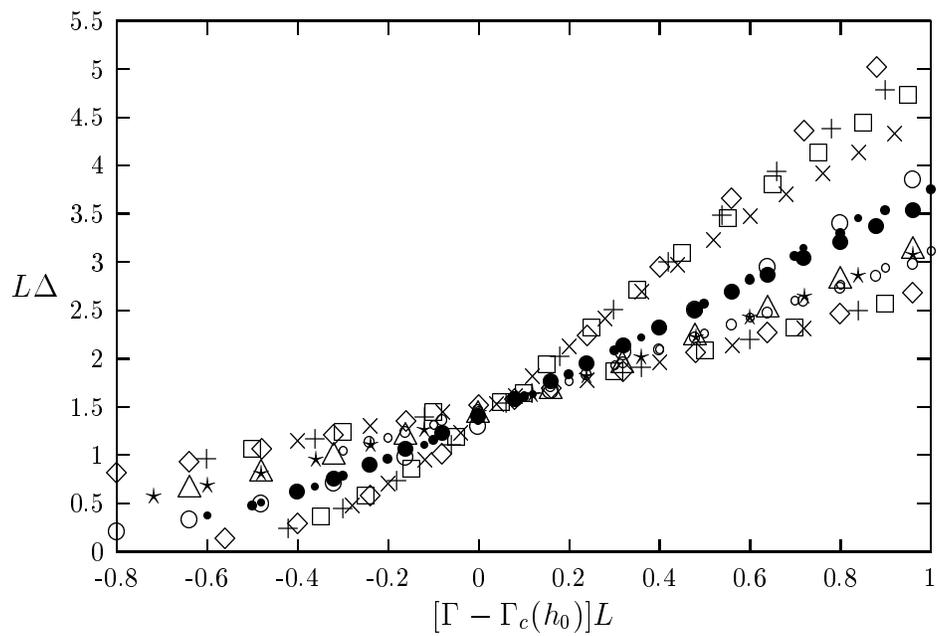
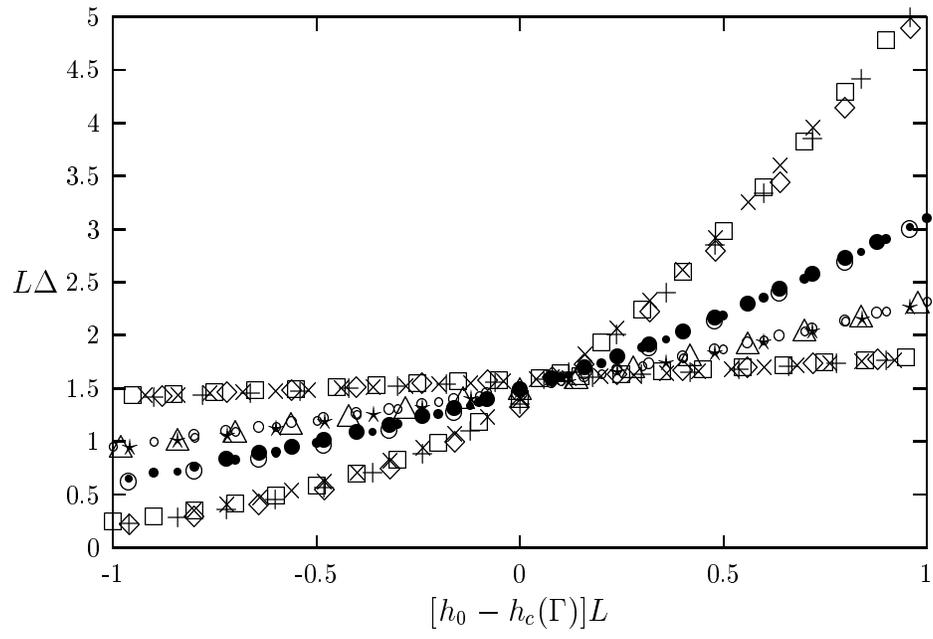
\end{document}